\documentclass{emulateapj}
\usepackage{amssymb,amsmath,epsfig,graphicx, natbib}

\def\sles{\lower2pt\hbox{$\buildrel {\scriptstyle <}
   \over {\scriptstyle\sim}$}} 
\def\sgreat{\lower2pt\hbox{$\buildrel {\scriptstyle >}
    \over {\scriptstyle\sim}$}} 

\def\kms{\mbox {~km~s$^{-1}$}}
\def\ergsec{\mbox {~erg~s$^{-1}$}}

\def\cmv{\mbox {~cm$^{-3}$}}

\def\mpc3{\mbox {~Mpc$^{3}$}}

\def\asec{\mbox {\ifmmode {'' }\else $''~$\fi}}  
\def\amin{\mbox {\ifmmode {' }\else $'~$\fi}}    
\def\arcsper{\mbox {\ifmmode \rlap.{'' }\else $\rlap{.}'' $\fi}} 
\def\arcmper{\mbox {\ifmmode \rlap.{' }\else $\rlap{.}' $\fi}} 
\def\SFR{\mbox {$\rm M_\odot ~yr^{-1}$}}

\def\Ha{\mbox {H$\alpha$}}
\def\Hb{\mbox {H$\beta$}}

\def\h0{\mbox {~H$_0$}}
\def\q0{\mbox {~q$_0$}}

\begin{document}

\title{The Emission-Line Spectra of Major Mergers: Evidence for Shocked Outflows
\footnote[1]{\lowercase{\uppercase{T}he data presented herein were obtained at the \uppercase{W.M. K}eck \uppercase{O}bservatory, which is operated as a scientific partnership among the \uppercase{C}alifornia \uppercase{I}nstitute of \uppercase{T}echnology, the \uppercase{U}niversity of \uppercase{C}alifornia and the \uppercase{N}ational \uppercase{A}eronautics and \uppercase{S}pace \uppercase{A}dministration. \uppercase{T}he \uppercase{O}bservatory was made possible by the generous financial support of the \uppercase{W.M. K}eck \uppercase{F}oundation.}}
}
\author{Kurt T. Soto$^1$, C. L. Martin$^1$, M. K. M. Prescott$^1$, L. Armus$^2$}
\affil{$^1$ Physics Department, University of California, Santa Barbara, CA 93106-9530 \\
$^2$ Spitzer Science Center, California Institute of Technology, Pasadena, CA.}

\begin{abstract}
\label{sect:abst}
Using a spectral decomposition technique (Soto \& Martin 2012, hereafter Paper~I), 
we investigate the physical origin of the 
high-velocity emission line gas
in a sample of 39 gas-rich, ultraluminous infrared galaxy (ULIRG) mergers.
Regions with shock-like excitation
were identified in two kinematically distinct regimes, characterized by broad 
($\sigma >$ 150 \kms) and narrow linewidths. Here we investigate the physical 
origin of the high-velocity (broad) emission with shock-like line ratios. 
Considering the large amount of extinction in these galaxies, the blueshift of 
the broad emission suggests an origin on the near side of the galaxy and 
therefore an interpretation as a galactic outflow. The large spatial extent 
of the broad, shocked emission component is generally inconsistent with an origin in 
the narrow-line region of a AGN, so we conclude that energy and momentum supplied
by the starburst drives these outflows. The new data are used to examine the
fraction of the supernova energy radiated by shocks and the mass loss rate
in the warm-ionized phase of the wind. 
We show that the shocks produced by galactic outflows can be recognized in moderately 
high-resolution, integrated spectra of these nearby, ultraluminous starbursts. The 
spectral fitting technique introduced in Paper~I may therefore be used to improve
the accuracy of the physical properties measured for high-redshift galaxies from
their (observed frame) infrared spectra.

\end{abstract}

\subjectheadings{galaxies: starburst --- galaxies: evolution --- galaxies: active --- 
galaxies: formation}

\section{Introduction}
\label{sect:intro}

Ultraluminous infrared galaxies (ULIRGs) are some of the most powerful galaxies in 
the local universe,
with $\rm{log(L_{IR}/L_{\odot})} >$ 12. Many ULIRGs exhibit disturbed 
morphology which indicates that they are major mergers \citep{Borne:1999p1037}. 
Tidal interactions in major mergers are believed to drive gas inflow that fuels both
the starbursts and the active galactic nuclei (AGN) in ULIRGs 
\citep{Toomre:1972p1069, Sanders:1986p1070, Springel:2005p41, Hopkins:2008p40}.

Models of major mergers also predict an early phase of supernova feedback followed by the emergence of an AGN
\citep{Springel:2005p41,Hopkins:2005p1073,Hopkins:2012p1217}. Starburst-driven outflows are found in 
 $75-80\%$ of ULIRGs \citep{Rupke:2002p1062,Martin:2005p24}.
Interstellar absorption lines in their spectra are blueshifted a few hundred \kms\ relative
to the systemic velocity set by the molecular gas and stars \citep{Armus:1990p976,Heckman:1990p710,
Martin:2005p24,Martin:2006p5,Rupke:2005p1029,Rupke:2005p1063,Rupke:2005p1028,Rupke:2005p1078}. 
AGN-driven outflows, in contrast, have only been found in the small subset of ULIRGs with Seyfert~1 
nuclei \citep{Rupke:2005p1029}, possibly indicating 
that these systems are no longer classified as ULIRGs by the time AGN drive the outflows. 
Because the measured stellar velocity dispersions of ULIRGs typically exceed the stellar 
rotation speed and the surface brightness profiles fit $r^{1/4}$ laws \citep{Genzel:2001p44}, 
as in elliptical rather than spiral galaxies, removal of the gas 
-- whether by star formation, black-hole fueling, or outflow --
provides the last step in transforming ULIRGs into field elliptical galaxies \citep{Dasyra:2006p234}.

The high SFRs common among $z \sim 2-3$ galaxies are thought to be fueled by steady gas accretion instead of
mergers \citep{Daddi:2007p1053, Shapiro:2008p990, Noguchi:1999p1220}, but mergers may play a pivotal 
role in the formation of galactic spheroids even at these redshifts \citep{Hopkins:2011p1232}. Regardless of how these galaxies
get their gas, comparable star formation rates are only found among ULIRGs and the Lyman-Break Analogs
\citep{Overzier:2008p1227,Overzier:2009p1228,Overzier:2009p1229,Overzier:2011p1230,Heckman:2011p1231}
in the local universe, so these environments provide the best local
laboratories for studying the feedback processes that shape the evolution of high-redshift galaxies.

Given the importance of incorporating feedback from massive stars and AGN into galaxy formation
simulations, simultaneously mapping out the excitation {\it and} gas kinematics in these extreme, local 
environments is of broad interest \citep{Heckman:1987p929,Armus:1990p976,Heckman:1990p710,Murray:2005p1072,Veilleux:2009p248}. The emission-line spectrum of shocked regions, for example, is easily distinguished 
from the spectrum of gas photoionized by massive stars but quite similar to the spectrum of gas 
photoionized by an AGN. Because the size of the narrow-line region (NLR) powered by AGN is limited by 
the AGN luminosity \citep{Bennert:2006p958,Bennert:2006p930, Greene:2011p943}, one strategy for breaking this degneneracy between excitation 
mechanisms is to resolve the location of the gas emitting the {\it shock-like} spectrum 
\citep[][, Paper~I]{MonrealIbero:2006p598, Rich:2011p1044, Goncalves:2010p1045}. Large velocity 
dispersions provide another way to identify regions with strong feedback from star formation and active 
nuclei, gravitational instability, and/or streams of recently accreted material \citep{Genzel:2011p1007, 
Law:2009p1011, ForsterSchreiber:2006p1033} and the Doppler shifts of broad emission-lines can be 
used to distinguish shocks generated by galactic winds and infall from either cold streams or 
tidally stripped gas. 

To provide a more comprehensive examination of the location and speeds of shocks in ULIRGs,
we mapped the emission-line ratios across 39 local ($z = 0.043 - 0.152$) ULIRGs in velocity 
{\it and} one spatial direction. 
These longslit spectra obtained with the Keck Echellette Spectrograph and Imager 
(ESI) lack the complete spatial coverage of integral field spectra (IFU) but offer more 
sensitivity and broader spectral coverage than IFU observations. In Paper~I, we identified
the regions with shock-like line ratios.
Following a brief summary of the sample and observations in Section \ref{sect:obs} of this
paper, the velocity dispersion and Doppler shift of the shocked regions are presented in 
Section \ref{sect:results}, where we argue that outflows are the 
primary origin of those shocked regions with broad ($\sigma > 150$\kms) emission lines.
In Section \ref{sect:outflow}, we discuss the origin of the outflows and estimate the outflow properties.
Finally, we discuss the impact of shocked gas on the integrated spectrum in Sect. \ref{sect:outflow_int}.

\section{Observations and Reductions}
\label{sect:obs}

This study examines the measurements presented in Paper~I of 39 
ULIRGs at various stages of merging chosen from from the IRAS 2~Jy survey \citep{Murphy:1996p23}. 
These local galaxies span $z = 0.043-0.163$, allowing spatially resolved optical 
spectroscopy with the Keck ESI 1\arcsec $\times$ 20\arcsec\ long slit. The data were reduced 
as described in \cite{Martin:2005p24}. Emission lines in the galaxies have $\sim$70 \kms\ 
spectral resolution and $\approx$ 0.8\arcsec\ spatial resolution limited by the atmospheric seeing. 
The median 
spatial resolution is 1.5 kpc, with a range of 0.7 to 2.1 kpc depending on redshift.

\section{Results}
\label{sect:results}

In Paper 1, we decomposed the emission line profiles
into multiple Gaussian components. An example of this
decomposition is shown in Figure 1. Because we jointly fit
several transitions, we measured diagnostic line ratios
for each velocity component. Using kpc-scale apertures,
we mapped the Doppler shift, velocity width, and line fluxes
across the ULIRGs. The results, provided in Table 3 of Paper 1, 
were used to classify the excitation mechanism of each
component as HII-like or shock-like. Here,
we explore the relationship between the gas kinematics and
the excitation mechanism.

\begin{figure}
\centering
\epsfig{file=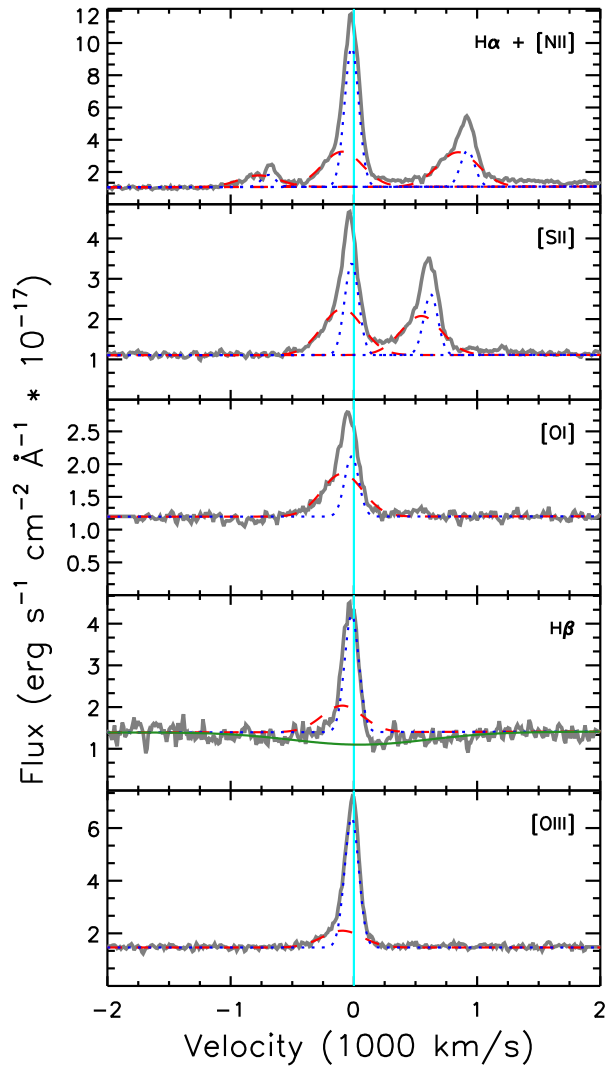, width=\linewidth, keepaspectratio=true}
\caption{\label{fig:trans_prof} 
Example of emission-line fit from paper I; the line profiles of the different
transtions were fit with common kinematic components of varying flux.
This figure demonstrates the clear variations in the line 
profiles for each transition, i.e. the blueshifted wing on \Ha, the nearly triangular line 
profile of [\ion{N}{2}]$\lambda$6583, and the very low intensity blue wing on [\ion{O}{3}]$\lambda$5007. 
Using this fitting method, we are able to deconstruct the line profile and examine the 
excitation mechanism for all of the spectral components that show the same kinematics.}
\end{figure}

For each velocity component at a given spatial location,
we classified the excitation mechanism as HII-like or shock-like
using diagnostic ratios of line fluxes \citep{Kewley:2006p38}.
The [OI]/Ha ratio plays a major role in this 
categorization due to its high sensitivity to shocks. 
In cases where [OI]/Ha is obscured by telluric absorption, 
we rely on [SII]/Ha and [NII]/Ha to make this distinction.
If the errors in the flux ratio cross the maximum star formation 
line, then the excitation mechanism is designated ``unclear''.
We note that our ``unclear'' category differs from the composite 
classification of \cite{Kewley:2006p38},
defined by line ratios
between the empirically defined maximum star formation limit
and the  maximum star formation.
We further describe these categories and the fitting method in
Paper I.

Figure 2 shows a histogram of the velocity dispersions of the components
from all apertures across all galaxies divided by excitation class.
Spectral components classified as ``HII-like'' are narrow in 
linewidth, having $\sigma_v \le 150$ \kms.
The components identified as ``shock-like'', however, 
present a  high-velocity tail at $\sigma_v > 150$ \kms.
Clearly the broad components are preferentially shock excited.

We show the Doppler shift of all the components in Figure \ref{fig:sig_vel}.
The narrow, shock-like components (and the HII components) are 
detected over distances of $15 - 30 $~kpc along the slit. The velocity offsets
relative to the galaxy redshift are positive in one direction and negative
to the opposite side of the nucleus. Figure~5 of Paper 1 demonstrates that 
the position-velocity diagrams for these narrow, shock-like components are typically 
consistent with the projection of a  galactic rotation curve, suggesting the process producing 
narrow, shock-like emission components is associated with a gas disk.

The broader components with  $\sigma_v > 150$~\kms, however, do not show similar
rotation gradients, suggesting that a different physical 
mechanism is involved in producing the broad emission.
Figure \ref{fig:sig_vel} shows that the broad, shocked 
components are nearly always blueshifted in stark constrast to the narrow components. 
The blueshifts of the broad, shocked components reach a maximum of 500~\kms. 

Because the ULIRGs are very dusty galaxies, their nuclei are not transparent at
optical wavelengths. 
Hence, the emission-line radiation that we detect at the 
center of a ULIRG must be emitted on the near side of the dusty, 
gaseous disk. 
We therefore conclude that the blueshifted emission comes from 
outflowing gas on the near-side of the galaxy rather than
infall on the far side. 
The broad emission in adjacent apertures maintain similar kinematics,
which allows us to assume that these regions have similar dust obscuration.
In support of this outflow interpretation,
we note that the broad, shocked emission spans a velocity range quite similar
to that of the blueshifted \ion{Na}{1} absorption troughs previously identified 
in these spectra \citep{Martin:2005p24, Martin:2006p5}.

Using the spatial information along the slit, we traced the extent of each
spectral component in Paper 1. Figure \ref{fig:pos_sig} shows how the spatial
extent of an emission component varies with its velocity disperion and excitation
classification. Broad, shocked emission in ULIRGs is usually spatially extended
but is not detected as far away from the nucleus as is the narrow component.
The broad, shocked component can typically be traced to $\sim 5$ kpc away from the peak 
continuum emission and up to 6 kpc in IRAS~03158+4227. We discuss the source of
these broad, shocked outflows in Section \ref{sect:outflow}.

\begin{figure}
\centering
\epsfig{file=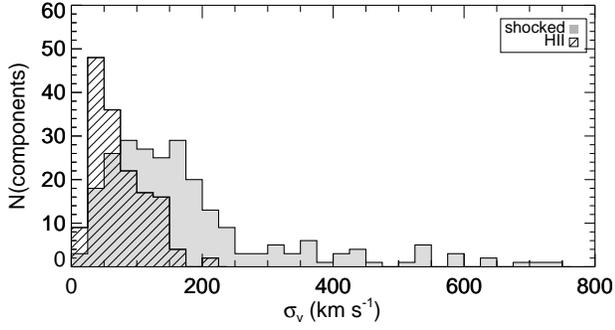, width=\linewidth, keepaspectratio=true}
\caption{\label{fig:hist} In the above histograms we present the distributions of 
line width for components identified as ``shock-like'' and those identified as ``HII''. Nearly 
all components with ``HII'' line ratios have $\sigma_v < 150 $\kms. The distribution of 
components identified as ``shock-like'' has a tail out to large line widths.
(Of the 430 components measured, 34 copomnents are not shown here because the measurement
uncertainties did not place them uniquely in one of these two excitation classes.)}
\vspace{0.4cm}
\end{figure}
\begin{figure}
\centering
\epsfig{file=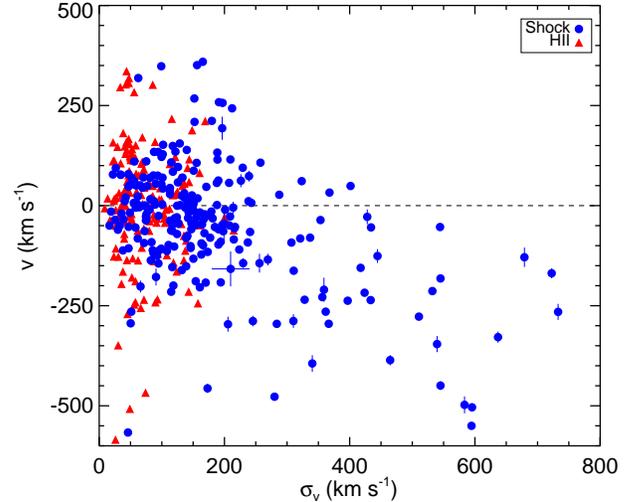, width=\linewidth, keepaspectratio=true}
\caption{\label{fig:sig_vel} We find that the components identified as shock-like with $
\sigma_v < 150$~\kms\ have a similar distribution in Doppler shift as the HII components. The 
shock-like components with $\sigma_v > 150$~\kms, however, present significant blueshifts.}
\vspace{0.4cm}
\end{figure}

\begin{figure}
\centering
\epsfig{file=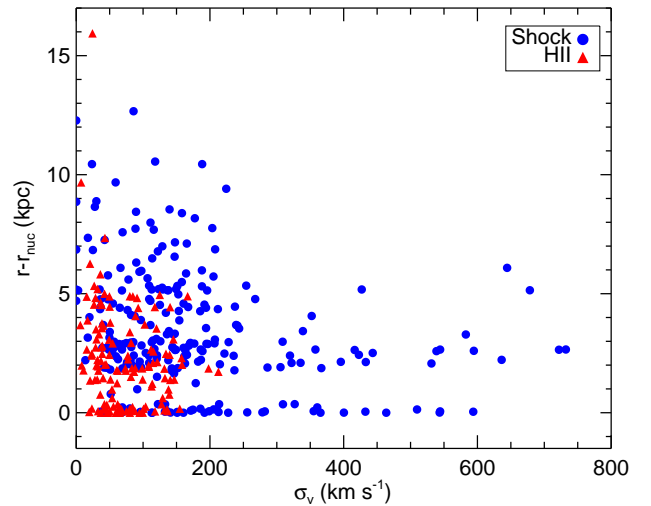, width=\linewidth, keepaspectratio=true}
\caption{\label{fig:pos_sig} HII and shock-like components with $\sigma_v < 150$~\kms\ extend 
to regions 15 kpc away from their associated nuclei. Components with $\sigma_v > 150$~\kms, 
however, remain within 5 kpc of the nucleus. The gap between the points at
${\rm r = r_{nuc}}$ and rest of the distribution reflects the spatial
resolution of the spectroscopy.}
\vspace{0.4cm}
\end{figure}

\section{Outflow Source}
\label{sect:outflow}

In general, the broad emission components are blueshifted to high velocity relative to the
systemic velocity and only a few components are redshifted as shown in Fig. \ref{fig:sig_vel}. 
For emission, a blueshifted line profile can be interpreted in two ways -- outflow 
on the near side of the galaxy or inflow on the far side of the galaxy. 
In the case of dusty ULIRGs, however, the presence of dust obscuration allows a
unique identification of the blueshift as outflow;
photons coming from infalling gas on the far side of the galaxy would have a longer path 
length through dusty regions, thereby increasing the amoung of extinction.
The outflow interpretation is further supported by the spatial extent of the blueshifted lines 
identified in Fig. \ref{fig:pos_sig}.
The spatial extent of broad blueshifted  emission suggests that the
emission is associated with regions beyond the deepest part of the 
gravitational potential well.
In the following sections, we explore the possible physical origins of the outflow.

\subsection{Star Formation vs. AGN as a Power Source}
\label{sect:agn_vs_sfr}

The power source of ULIRGs has been under debate for more than 20 years, owing to the obscuration 
in optical and shorter wavelengths by dust. 
Using the Spitzer Infrared Spectrograph (IRS), \cite{Veilleux:2009p248} examined the power 
source of 14 of the 39 ULIRGs included in the Paper I sample by measuring indicators of 
AGN contribution from infrared indicators.
Indicators of the strength of AGN contributions come from ratios of mid infrared emission
lines
[\ion{O}{4}]/[\ion{Ne}{2}] and [\ion{Ne}{5}]/[\ion{Ne}{2}]
\citep{Armus:2004p1195,Armus:2006p1196,Armus:2007p947,Veilleux:2009p248},
however, these ratios are well constrained for only 3 of the 14 ULIRGs in this sample. 
Two of the objects with well constrained contributions (IRAS01572+0009, IRAS12072-0444)
show $\sim 90 \%$ AGN contributions to L$_{\rm bol}$.
The third object (IRAS15001+1433) that is well constrained shows an 
$\sim 50\%$ AGN contribution.
Measurements of the mid-IR line ratios for the other 11 ULIRGs place a $\sles\ 30\%$ upper limit
on the AGN contribution to their bolometric luminosity. Although the AGN is generally
not the dominant source of luminosity in these ULIRGs, we still need to investigate
whether the broad lines could be directly related to the AGN.

\subsubsection{AGN Broad Line Region}
\label{sect:blr}

A broad line region (BLR) is a common feature in Seyfert 1 and 1.5 galaxies, appearing as permitted 
and intercombination emission lines broadened to 1,000 to 25,000 \kms\ FWHM
\citep{Osterbrock:2006p1109}.
The weakness of forbidden transitions in the BLR is explained by the high
electron density, $n_e > 10^9 \cmv$, which enhances collisional de-excitation
\citep{Peterson:2006p1138}. 
Reverberation mapping places the BLR size at much less than 1~pc 
\citep{Bentz:2009p1041,Brewer:2011p1040}, suggesting fast moving gas reflects the deep
gravitational potential in the vicinity of a supermassive black hole \citep{Dietrich:1999p1141}.
In our ULIRG spectra, emission from an AGN BLR would be confined to
 the central aperture~($\sim 1\arcsec$). 

In the ULIRG spectra, however, the broad emission lines extend beyond a spatial
resolution element,  calling into question the BLR as a source of these features. 
Additionally the broad emission features appear not only in the permitted transitions, 
but also in the forbidden transitions. 
These observations, along with the low electron density measurements $\sles\ $ 100 \cmv\
from [\ion{S}{2}] line ratios, rule out an AGN BLR as the source of the broad, shocked
emission components.

\subsubsection{Narrow Line Region}
\label{sect:nlr}

In the emission-line ratio diagrams, we use the line defining the maximum excitation 
	via extreme star formation \citep{Kewley:2006p38} to distinguish HII emission 
	from shock-like emission.
	Above this, line ratios are typically attributed to narrow line regions of Seyfert 
		galaxies or LINERs. 
	The Seyfert and LINER line ratios can be described by fast shocks where the gas 
		density ahead of the shock determines the influence of the radiative precursor
		on the overall emission line ratio.
	Gas rich Seyfert galaxies have a larger contribution from the radiative precursor, 
		resulting in larger [\ion{O}{3}]/\Hb\ ratios.
	LINERs are suspected to have lower gas densities ahead of the shock, which leads to 
		lower [\ion{O}{3}]/\Hb\ ratios \citep{Dopita:1995p1120}.

The size of shocked AGN NLR region is empirically 
	related to the luminosity of the [\ion{O}{3}] $\lambda 5007$ emission 
	\citep{Bennert:2002p907,Greene:2011p943}.
	The scaling relation we use \citep{Greene:2011p943} is developed without corrections for 
		in situ dust extinction of L$_{[\rm OIII]}$, owing to the difficulty of robust 
		corrections for obscured quasars \citep{Reyes:2008p1135}.		
	The [\ion{O}{3}] luminosities (L$_{[\rm OIII]}$) used in this scaling relation, 
		only includes spectral components where log([\ion{O}{3}]/\Hb) is 
		greater than 0.5 to avoid flux contributions from an HII region.
	For each galaxy, we estimate  L$_{[\rm O III]} $ from the broad, shock-like 
		components and predict the expected size of an AGN NLR.

Since any nucleus can host an AGN, we measure luminosity for each galaxy in the 
	pairs of merging galaxies -- increasing the total galaxy count in this study to 48.
	Of the galaxies with components that have log([\ion{O}{3}]/\Hb) $>$ 0.5, 
		the range of luminosities spans several orders of magnitude 
		($1.9\times10^{37} < {\rm L_{[OIII]}} < 3.2\times10^{42}$ \ergsec) 
		with a median ${\rm L_{[OIII]}} = 8\times10^{39}$ \ergsec.
	For 9 galaxies (in 7 ULIRGs -- IRAS00188-0586, IRAS01003-2238, IRAS05246+0103, 
		IRAS08311-2459, IRAS09583+4714, IRAS13451+1232, and IRAS15130-1958) 
		the measured outflow radii are within a factor of 1.6
		of inferred radii from the scaling relation.

The majority of galaxies have L$_{\rm [OIII]}$ and spatial distributions
	that do not suggest excitation via AGN. 
	In the 8 ULIRGs (13 galaxies), no components appear above the log([\ion{O}{3}]/\Hb) $>$ 0.5 
		cutoff, but we still detect shock-like line ratios up to 6 kpc from the closest nucleus,
		suggesting some other mechanism is responsible for the ionization. 
	In 14 of these galaxies the components with appropriate log([OIII]/\Hb) 
		appear only in regions outside the nuclear aperture.
	Galaxies that have a strong extended AGN presence would more likely consist of a 
		contiguous region that includes the nucleus, suggesting that these are also not 
		part of an NLR.
	The remaining galaxies have regions with shocked gas at radii 
		greater than the estimated NLR size by a factor of 2.

In general, most cases do not favor AGN as the source of the observed emission line profile and 
instead are consistent with shocks driven by stellar winds and supernovae, as indicated by the AGN fractions. 
With this understanding of the outflow source, we can make estimates of the 
outflowing masses and the energy injected into the system by these processes.

\begin{deluxetable*}{lcccccccccccc}
\tablewidth{\linewidth}
\tablecaption{Outflow Parameters \label{tab:outflow}}
\tabletypesize{\scriptsize}
\tablehead{
IRAS Name &                 
L$_{\rm IR}$ &              
v$_{\rm out}$ &             
r$_{\rm out}$ &             
t$_{\rm out}$ &             
L$_{\rm tot}$ &             
M$_{\rm L}$ &               
M$_{\Sigma}$ &              
$\dot{\rm M}_{\rm L}$ &     
$\dot{\rm M}_{\Sigma}$ &    
SFR &                       
$\eta_{\rm L}$ &            
$\eta_{\Sigma}$ \\        
(1)&(2)&(3)&(4)&(5)&(6)&(7)&(8)&(9)&(10)&(11)&(12)&(13)
}
\startdata

IRAS00153+5454  &    12.10  &      270  &      3.71  &    13.31  &     51.5  &     17    &     47   &     1.3 &     3.6 &       68  &    0.02  &    0.05  \\  
IRAS00188-0856  &    12.33  &      660  &      1.57  &     2.32  &     407.  &     130   &     31   &     58  &     13  &      128  &    0.45  &    0.10  \\  
IRAS01003-2238  &    12.25  &     1140  &      3.66  &     3.13  &     203.  &     67    &     155  &     21  &     50  &       96  &    0.22  &    0.52  \\  
IRAS01298-0744  &    12.29  &      240  &     11.27  &    44.98  &     86.6  &     28    &     620  &    0.6  &     14  &       57  &    0.01  &    0.24  \\  
IRAS03158+4227  &    12.55  &     1330  &      7.22  &     5.30  &     181.  &     59    &     423  &     11  &     80  &      191  &    0.06  &    0.42  \\  
IRAS05246+0103  &    12.05  &      250  &      2.71  &    10.54  &     33.1  &     10    &     22   &     1   &     2.1 &       60  &    0.02  &    0.04  \\  
IRAS05246+0103  &    12.05  &      850  &      3.10  &     3.56  &     520.  &     170   &     110  &     48  &     31  &       60  &    0.80  &    0.52  \\  
IRAS08311-2459  &    12.40  &      600  &      4.94  &     8.08  &     338.  &     110   &     320  &     14  &     40  &      135  &    0.10  &    0.30  \\  
IRAS09111-1007  &    11.98  &      290  &      1.38  &     4.68  &     26.3  &     8     &     7.3  &     1.8 &     1.6 &       51  &    0.04  &    0.03  \\  
IRAS09583+4714  &    11.98  &      350  &      4.42  &    12.27  &     413.  &     140   &     320  &     11  &     26  &       51  &    0.20  &    0.50  \\  
IRAS10378+1109  &    12.23  &     1000  &      3.80  &     3.72  &     267.  &     88    &     150  &     24  &     41  &       92  &    0.26  &    0.45  \\  
IRAS10494+4424  &    12.15  &      730  &      4.34  &     5.78  &     80.2  &     26    &     88   &     4.6 &     15  &       76  &    0.06  &    0.20  \\  
IRAS10565+2448  &    11.98  &      250  &      2.66  &    10.54  &     23.2  &     8     &     29   &    0.7  &     2.8 &       66  &    0.01  &    0.04  \\  
IRAS11095-0238  &    12.20  &      260  &      3.65  &    13.61  &     145.  &     48    &     120  &     3.5 &     8.8 &       95  &    0.04  &    0.09  \\  
IRAS12071-0444  &    12.31  &      570  &      7.95  &    13.57  &     15.9  &     5     &     210  &    0.4  &     15  &       25  &    0.02  &    0.61  \\  
IRAS15130-1958  &    12.03  &      890  &      3.65  &     4.03  &     29.5  &     10    &     51   &     2.4 &     13  &       24  &    0.10  &    0.53  \\  
IRAS16090-0139  &    12.48  &      640  &      6.61  &    10.07  &     17.0  &     6     &     86   &    0.6  &     8.5 &      179  &    0.003 &    0.05  \\  
IRAS16487+5447  &    12.12  &      400  &      5.79  &    14.01  &     147.  &     49    &     360  &     3.5 &     26  &       71  &    0.05  &    0.36  \\  
IRAS17028+5817  &    12.11  &      200  &      1.33  &     6.42  &     121.  &     40    &     14   &     6.2 &     2.1 &       69  &    0.09  &    0.03  \\  
IRAS18368+3549  &    12.19  &      370  &      4.09  &    10.70  &     333.  &     110   &     140  &     10. &     13  &       83  &    0.12  &    0.16  \\  
IRAS19297-0406  &    12.36  &      730  &      4.30  &     5.74  &     233.  &     77    &     180  &     13  &     31  &      123  &    0.11  &    0.25  \\  
IRAS20087-0308  &    12.39  &      330  &      5.62  &    16.48  &     1170  &     390   &     760  &     24  &     46  &      132  &    0.18  &    0.35  \\  
IRAS23327+2913  &    12.03  &      630  &      3.81  &     5.95  &     60.5  &     20    &     99   &     3.4 &     17  &       58  &    0.06  &    0.29  \\  
IRAS23365+3604  &    12.13  &      580  &      0.89  &     1.50  &     5.52  &     2     &     1.6  &     1.2 &     1.1 &       73  &    0.02  &    0.02  \\  

\enddata
\tablecomments{ 
Column 1: IRAS Name.
Column 2: Total infrared luminosity log(L$_{\rm IR}$/L$_\odot$) \citep{Murphy:1996p23}.
Column 3: Outflow velocity taken as $v_{\rm out} = \left|v\right| + \sigma_{v}$ (\kms) from 
the Paper I (Table 3). Since we are unsure of the degree to which projection effects play a 
role in the measured velocity, assume all outflowing regions have the same $v_{\rm out}$, but 
is foreshortened. In this simplification we take the maximum $v_{\rm out}$ of the apertures, 
to represent the vector that is most parallel to the line of sight.
Column 4: Maximum radius of outflow (kpc).  $r_{out}$ is defined by the 
maximum separation between an aperture which shows outflow and the continuum peak. 
Column 5: outflow time scale (Myr).
Column 6: Total luminosity of \Ha\ from the outflow components (L$_{\rm tot}/10^{40}$) erg~s$^{-1}$.
We correct the \Ha\ emission for Galactic dust extinction using the reddening
curve from \cite{Cardelli:1989p27}  and values of E(B-V) and A$_V$ from the 
Infrared Science Archive maps \citep{Schlegel:1998p978}. 
We use the measured Balmer decrement and the \cite{Calzetti:1994p934} reddening curve to correct
for internal extinction.
Column 7: Mass of outflowing gas in $10^6$ M$_\odot$ from L$_{\rm tot}$ and Eq.\ref{eq:m_lum} 
assuming $n_e = 100 {\rm \cmv}$ and $\gamma' = \gamma_{phot}$.
Column 8: Mass of outflowing gas in $10^6$ M$_\odot$ from $\Sigma$ and Eq.\ref{eq:m_surf} with an 
assumed hemisphere geometry, with the maximum path length through the sphere defined by the 
radius, $\delta s = r_{out}$, and $f =0.01$. Within each aperture the path length through the sphere varies as $s 
= (r_{\rm out}^2 - (r_{\rm aperture} - r_{\rm nucleus})^2) ^{1/2}$
Column 9: Outflow rate estimated from L$_{\rm tot}$ in M$_\odot$~yr$^{-1}$ described in Section \ref{sect:vel_time}. 
Column 10: Outflow rate estimated from $\Sigma$ in M$_\odot$~yr$^{-1}$ with an assumed 
hemisphere geometry. 
Column 11: Star formation rate estimated from L$_\mathrm{IR}$ are values \citep{Murphy:1996p23}
 and relation SFR = L$_\mathrm{IR}/13.0 \times 10^9 $L$_\odot$. This relation is an 
adjustment of the \cite{Kennicutt:1998p3} relation adjusted for a stellar mass range from 1 
to 100 M$_\odot$ (M$_\odot~{\rm yr}^{-1}$). This star formation rate is also corrected for 
AGN contribution to L$_{\rm bol}$ for the objects included in \cite{Veilleux:2009p248}.
In cases where upper limits to the AGN fraction are not provided for an individual object in 
\cite{Veilleux:2009p248}, we use the average AGN contribution of 30\%.
Column 12: Efficiency of outflow from $\dot{\rm M}_{\rm L}$.
Column 13: Efficiency of outflow from $\dot{\rm M}_{\Sigma}$ with an assumed hemisphere 
geometry and filling factor $f = 0.01$.
The objects in this table are restricted to those that have at least 1 aperture with a shock-
like spectral component with $\sigma_v > 150$ \kms.}
\end{deluxetable*}

\subsection{Energetics of Supernova Feedback}

Excluding the AGN-dominanted cases IRAS01572+0009 and IRAS12072-0444, 
the star formation rates in these ULIRGs indicate enormous amounts of
mechanical power is deposited in the ISM by massive stars. In this section,
we compare mechanical power to the power in the bulk outflow of warm-ionized gas.  
We stress that our estimates of the mass and kinetic energy in the bulk flow
are no better than factor of $\sim 3$ accuracy. Uncertainty about the filling 
factor of the warm-ionized gas and the large-scale geometry of the outflow
dominate the systematic error. The mass, energy, and momentum carried by the
outflowing, warm-ionized gas should clearly be viewed as a lower limit on
the total amounts carried by all phases of the wind, which likely includes
substantial components of both coronal gas and molecular gas.

\subsubsection{Mass of Warm-Ionized Gas} \label{sect:mass}

For each ULIRG that shows broad, shock-like emission components,
we can estimate the outflowing mass of warm-ionized gas. This mass is 
the product of the gas density and the volume occupied by the warm-ionized
phase. 

The fraction of the total volume V filled with
warm-ionized gas is called the filling factor. For HII regions, this
filling factor,
\begin{equation} \label{eq:f}
f \equiv \frac{\left<n_e^2\right>}{n_e^2},
\end{equation}
is computed from the RMS electron density obtained from the \Ha\ luminosity or
surface brightness and the electron density computed from a density-sensitive
doublet ratio such as [\ion{S}{2}] $\lambda \lambda 6717, 31$.
Values of the filling factor range from $0.001 < f < 0.1$ in nearby HII regions
\citep{Searle:1971p1083,Kennicutt:1984p1084,Kaufman:1987p1086} and therefore have
a very signficant impact on the estimated mass. 
The range of filling factor may indeed be different in the ULIRGs, given the 
difference in star formation rate from these HII regions in normal galaxies, but these 
values provide a point of comparison.
In the ULIRG spectra, the flux ratio for the [\ion{S}{2}] doublet is typically
consistent with the low-density limit and indicates $n_e~\sles~100\cmv$, 
but the ratios of the two [\ion{S}{2}] lines indicate densities $n_e~\approx~500\cmv$ 
in the central apertures  of IRAS01003-2238, IRAS05246+0103, IRAS08311-2459, IRAS09583+4714, 
and IRAS23327+2913. For our mass estimates, we adopt a fiducial value of $n_e = 100 \cmv$
but carry along the scaling with density explicitly in our results.

The outflow mass is
\begin{equation} \label{eq:m_geom}
M_{out} = \mu m_{\rm H} {\rm V} n_e f,
\end{equation}
where $n_e$ is the electron density, $\rm m_H$ is the mass of hydrogen, and the mass per H atom is
$\mu = 1.4$ amu when helium is included. The unknown factor
$V n_e f$ in Eqn. \ref{eq:m_geom} is simply related to the \Ha\ luminosity
by
\begin{equation} \label{eq:L_phot}
{\rm L_{H\alpha}} = \gamma' \left<n_e^2\right> {\rm V} = \gamma' n_e^2 f {\rm V}.
\end{equation}
Eliminating the common factor between Eqn. 2 and 3, the outflowing mass,
\begin{equation} \label{eq:m_lum}
M_{\rm L} = \frac{\rm \mu m_H L_{H\alpha}}{\gamma' n_e},
\end{equation}
can be estimated from the luminosity of the broad, shocked emission component 
for any assumed value of the gas density, $n_e$.

The effective volume emissivity, $\gamma'$, varies depending on the temperature and excitation
mechanism. For photoionized regions with T = $10^4$ K, 
$\rm \gamma' = \gamma_{\rm phot} = \alpha_{\rm H\alpha}^{eff} h \nu = 3.56\times 10^{-25}~erg~cm^{3}~s^{-1}$
in Case~B recombination theory \citep{Osterbrock:2006p1109}. 
In shocked regions,  collisional excitation and ionization can make $\gamma'$ larger or smaller than 
$\gamma_{\rm phot}$ by a factor of two to three \citep{Goerdt:2010p1234,Genzel:2011p1007}. 
Since the photionization case is in the middle of the plausible range for $\gamma_{\rm phot}$, we
estimate masses using $\gamma_{\rm phot}$.

\begin{figure}[h]
\centering
\epsfig{file=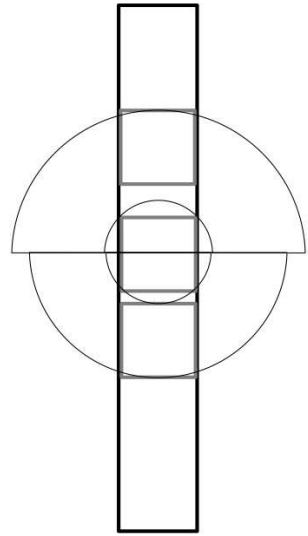, width=0.75\linewidth, keepaspectratio=true, trim = 20 70 20 110 , clip=true}
\caption{\label{fig:circ_est} The long black rectangle represents the slit, 
the gray boxes represent apertures along the slit, the annuli are the total 
estimated regions for each position. 
The number of outflows detected in ULIRGs \citep[75-80\%][]{Rupke:2002p1062,Martin:2005p24} 
suggests that a biconical outflow cone has a large opening angle $\sim$145\degr.
This allows us to use a sphere as a rough approximation to the geometry.
We estimate the luminosity contributed by regions outside of the slit by assuming this circular 
symmetry and a constant surface brightness within the half annuli as the aperture within it. 
The central aperture coincides with the nucleus of the galaxy used in the measurement.}
\end{figure}

We present the estimated masses in Table \ref{tab:outflow}. They range
from $\rm 1.8 \times 10^{6}~M_{\odot}$ to $\rm 3.9 \times 10^{8}~M_{\odot}$. 
The median and standard deviations are $\rm 40 \times 10^{6}~M_{\odot}$ and  
$\rm 84 \times 10^{6}~M_{\odot}$ respectively. 
The systematic errors in the mass estimate are difficult to quantify. For example,  
Fig.~\ref{fig:circ_est} illustrates the aperture correction to the
observed flux. 
In addition, corrections for internal \Ha\ extinction can be large. The large outflow
mass in IRAS20087-0308, for example, is due in large part to the extinction correction.
For individual galaxies, the masses in Col.~7 of Table~1 should be interpreted as
rough estimates accurate to a factor of 2-3.

The measured \Ha\ surface brightness provides a sanity check on the assumed electron density,
or equivalently $\left<n_e^2\right>^{1/2} f^{-1/2}$ by Eqn.~\ref{eq:f}.
The \Ha\ surface brightness, $\Sigma = L_{Ha} / A $ depends on the three-dimensional shape
of the outflow since the volume is the product of the projected area $A$ and the depth
of the emitting region, $\Delta s$, along the sightline. 
The surface brightness depends on density as
\begin{equation} \label{eq:ne_surf}
\Sigma =  \left<n_e^2\right> \gamma' \Delta s
\end{equation}
By analogy to Eqn. 4, the inferred outflow mass is
\begin{equation} \label{eq:m_surf}
M_{\rm \Sigma} = \left( \frac{\mu m_H A \Sigma}{\gamma' n_e} \right) = 
\mu m_H L  \left(\frac{\Delta s f}{\gamma' \Sigma}\right)^{\frac{1}{2}}.
\end{equation}
In Table 1, col.~8 provides examples of the mass estimates obtained by this method.

If our density estimate $n_e$ is reasonable, then equating $M_{\Sigma}$ and
$M_{L}$ should imply reasonable values for  the product of the filling factor and 
the depth of the emitting region along the line of sight. 
If we adopt $f=0.01$, then the required path length through the emitting 
region is often only a few hundred pc. Such a small pathlength would arise only if the
emitting region had a hollow shell structure. If the emitting region is roughly
spherical, $\Delta s \approx \sqrt{A}$, the implied filling factor is often small,
 $f \sim 10^{-3}$. The real values of $s$ and $f$ that unify the mass estimates 
may involve a change in both of these parameters, but will depend on each individual case.
The bottom line is the volume filling factor of the warm-ionized gas may be tiny thereby
leaving lots of volume to be filled by gas at a different temperature.

\subsubsection{Outflow Velocity and Time Scale}
\label{sect:vel_time}

The time scale for the outflow is estimated from the geometry and the kinematics measured from the emission
line profile.
In Table \ref{tab:outflow}, we define the outflow velocity ($v_{\rm out}$) as 
$\left|v\right| + \sigma_{v}$ (\kms) taken from Table 3 in Paper I.
The median $v_{\rm out}$ in this sample is 580 \kms, and the values range from 200 to 1330 \kms.

Estimating the outflow timescale from the outflow radius and velocity, $t_{out} = r_{out}/v_{out}$,
the implied mass loss rate of warm-ionized gas is $\dot{\rm M} = {\rm M}_\mathrm{out} / \tau$. 
For the estimate of mass outflow using luminosity ($\rm M_L$), we find a range from 0.4 to 58 
\SFR\ where the median is 4.6 \SFR. 
The largest $\dot{\rm M}$ is in IRAS00188-0856, due mostly to the moderate 
v$_{\rm out}$ and small r$_{\rm out}$. 
The surface brightness mass estimate with a spherical geometry ($\rm M_\Sigma$) gives larger outflow rates on 
average from 1 to 80 \SFR\ where the median is 15.2 \SFR. 
In this case the largest outflow rate is in IRAS03158+4227, 
which has a large v$_{\rm out}$ and r$_{\rm out}$, but in the surface brightness case, 
the large r$_{\rm out}$ has a greater influence.

\subsubsection{$\dot{M}$ vs SFR}
\label{sect:mdot_vs_sfr}

An important parameter in modeling galactic winds is the efficiency ($\eta$) that star forming regions 
are able to remove gas from the star forming region. 
This parameter informs the timescale over which a galaxy will form stars before the stellar 
population itself removes gas and shuts down star formation. 
We estimate $\eta$ as the ratio of outflow mass to the star formation rate measured from 
IR luminosities in \cite{Murphy:1996p23} (Table~\ref{tab:outflow}).
We use the star formation rate estimated from the SFR = L$_\mathrm{IR}/13.0 \times 10^9 $L$_\odot$ 
\citep{Kennicutt:1998p3}, adjusted for a stellar mass range from 
1 to 100 M$_\odot$ and correcting the L$_{\rm IR}$ for the AGN fraction.

In this sample of galaxies, the outflow rates  typically yield
$\eta \sim 10^{-1}$ but span the range $10^{-3} < \eta < 1$. We emphasize, however,
that this measurement refers only to the warm ionized gas, which
may be the dominant phase of the outflowing gas. 
The mass outflow rates in the cool gas traced by Na~I are
at least a few tenths of the star-formation rate, $\eta \equiv \dot{M} / SFR \approx 0.1$,
but are poorly constrained due to the large variations in ionization parameter among 
ULIRGs \citep{Martin:2005p24,Murray:2007p1042}. Hot winds and molecular outflows
may also carry significant mass.

\subsubsection{Shock Energetics}
\label{sect:shock_energy}

The presence of outflows implies that mechanical energy is injected into the gas. For objects that
show outflow in emission, we can estimate the energy injection rate into the gas from a starburst by using
the star formation rate.
We scale the predicted supernova energy injection rates from \cite[][SB99]{Leitherer:1999p1117}
to the calculated SFR for each source, after correcting for differences in the assumed IMF 
low mass cutoff (Kennicutt 0.1 - 100 M$_\odot$, SB99 1-100 M$_\odot$).
We assume continuous star formation model over a timescale longer than 40 Myr.
This produces a  nearly continuous injection of feedback energy contributed by supernovae
and stellar winds.

The simplest model for the initial response of the interstellar gas to this feedback
is based on stellar wind bubbles \citep{Weaver:1977p1115,Shull:1979p1119}. Shocks
direct 55\% of the feedback energy into the thermal energy of low density, coronal
bubble and the other 45\% into an expanding shell of swept-up interstellar medium.
About 60\% of the shell energy is radiated away in this model, suggesting that luminosity
of the shocks would be 27\% of the feedback power. 
For typical shock velocities, most
of this energy will come out in ultraviolet and optical emission lines. 
The energy in the optical lines will be slightly greater than the UV lines by $\sim 20\%$, 
suggesting that $\sim 15\%$ of the feedback power will come out in optical emission
\citep{Shull:1979p1119}.

\begin{figure}[h]
\centering
\epsfig{file=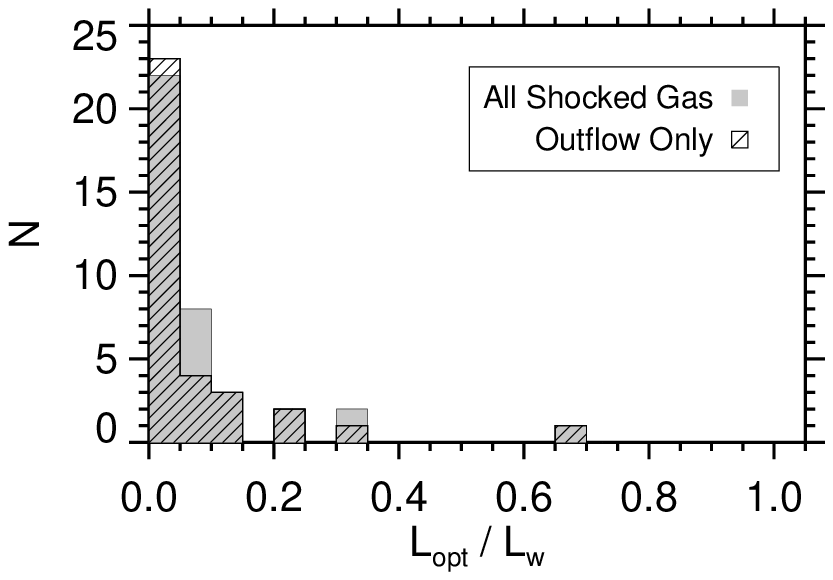, width=\linewidth, keepaspectratio=true}
\epsfig{file=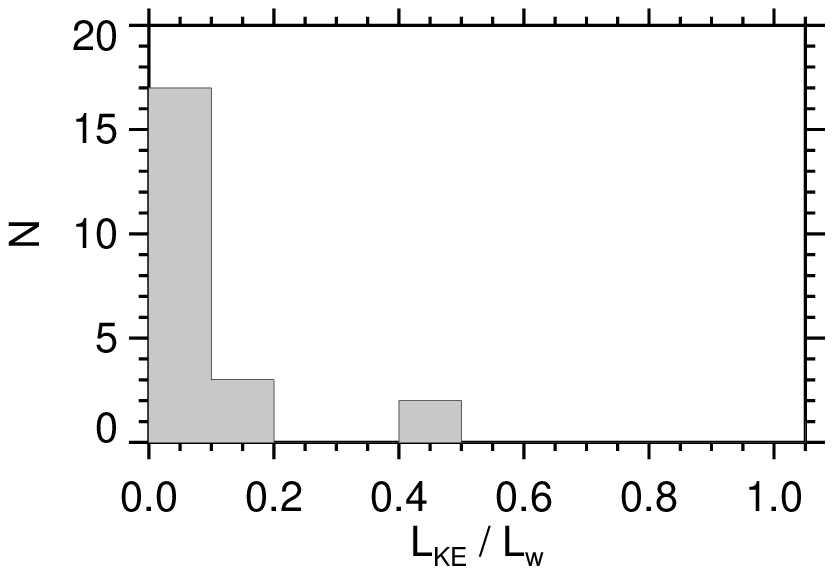, width=\linewidth, keepaspectratio=true}
\caption{\label{fig:nrg} {\it above:} A histogram showing the measured luminosity ($\rm L_{opt}$) emitted in shocked 
gas compared to the mechanical energy injection rate ($\rm L_w$) from supernovae scaled from the IR
star formation rate with SB99. For a whole galaxy, both the emission associated with the 
shocked outflow (hashed) and the emission from all shocked regions (solid) are generally 
in the range 0 - 15\%, with the peak below 5\%. The measured values are consistent with 
\cite{Weaver:1977p1115}. {\it below:} The fraction of the feedback energy that
contributes the kinetic energy in the mass outflow is again consistent with the Weaver model.
}
\end{figure}

Using our estimates of physical properties in Section \ref{sect:mass} and the shock speeds from
Paper I, we can make comparisons to these efficiencies.
In Figure \ref{fig:nrg} we show the ratio of the total optical luminosity ($\rm L_{opt}$) for the components indicating 
shocked outflowing gas in the measured lines compared to the mechanical power ($\rm L_w$, inferred from the SFR).
Since the luminosity of the optical emission lines is generally less than $\sim$15\%, it follows that
supernova feedback is a plausible source of power for this shocked emission.

Luminosity in the optical emission lines for 4 of the galaxies (IRAS05246+0103, IRAS09583+4714, IRAS13451+1232, and IRAS20087-0308)
exceeds the expected fraction of energy injected into outflow by supernovae.
The excess emission line energy in these objects (except for IRAS20087-0308), may be due to 
additional emission from an AGN NLR, since for these objects, the emission region size 
from scaling relations is consistent with that of an AGN NLR.
Emission from IRAS20087-0308 encounters heavy extinction making the Balmer decrement
difficult to measure, resulting in a possible over correction to the extinction, which would 
then skew the L$_{\rm opt}$/L$_{\rm w}$ ratio.

We calculate the ratio of power that appears as kinetic energy from the outflow rate ($\rm L_{KE}$) to 
the mechanical energy injection rate from the estimated mass and velocities measured for
the outflow regions in Table \ref{tab:outflow}.
This kinetic energy estimate uses $\dot{\rm M}_{\rm L}$ as the outflow mass estimate and $v_{\rm out}$ 
for the velocity.
In Figure \ref{fig:nrg}, we show that this energy rate is again consistent with the fraction of energy in
outflow models (mean = 0.08), which predict a $\sim 20\%$ fraction of 
mechanical energy in the kinetic energy of the shell.
The consistency in kinetic energy fraction with the models suggests that this process 
is possibly at work driving the outflows in ULIRGs. It also implies that the estimated masses are correct 
within an order of magnitude.

\section{Discussion}
\label{sect:discussion}

\subsection{Outflows in Integrated Spectra}
\label{sect:outflow_int}

In observations of galaxies at high redshift, obtaining similar spatial resolution to this 
study is difficult due to the small angular size of galaxies.
 The degree to which spatially integrated emission lines show evidence for the 
outflows in these spectra is unclear. 
 We examine spatially integrated spectra in these data to examine which of the 
resolved features remain evident in the integrated spectra.

\subsubsection{Spectral Classifications from the Total Line Flux}
\label{sect:tot_sclass}

In low spectral resolution, spatially unresolved studies of galaxies, one means of examining the 
excitation mechanism is through total line flux classifications.
We apply the same classifications (Paper I) to the ULIRGs based on spectra 
obtained by integrating along the slit.
 These total line flux classifications may be influenced by the observed shocks in
outflows and gas disks.  
 We examine the relation between classification and the presence of these shocks 
to determine which of the Seyferts, LINERS, and HII classified galaxies
host shocked gas.

All 6 galaxies classified as Seyferts present broad, blueshifted outflows in the kpc scale
apertures. 
 The narrow emission line components in these Seyferts also present shock-like ratios
in four out of six cases -- IRAS00188-0856, IRAS01003-2238, IRAS13451+1232, and IRAS15130-1958.
 These four Seyfert galaxies have both narrow and broad emission components that 
present shock-like flux ratios. The luminosity of the broad, shocked component
is sufficient L$_{\rm [O III]}$ to generate a spatially resovled NLR.
We conclude AGN photoionization is the source of
excitation for these four galaxies.

IRAS09583+4714 and IRAS12071-0444 present the two exceptions for the Seyfert classified 
galaxies, where the broad components differ from the narrow components in their 
excitation. 
 Of the two galaxies in the double IRAS09583+4714, the galaxy with shocked outflow 
presents narrow component  excitation consistent with HII and has a strong, 
disk-like rotation gradient. 
 In IRAS12071-0444, the narrow components have excitation more consistent with a LINER
classification and are highly extended to $\pm 9$ kpc.
 For both of these exceptional galaxies, scaling relations from L$_{\rm [O III]}$ 
suggest that the outflow region is larger than what would be expected from 
an AGN NLR.
 The inconsistency of the outflow excitation with the integrated classification in
these two galaxies suggest that a significant fraction of ULIRGs may be 
misclassified as Seyfert galaxies, when they host shocks in outflows outflows
driven by star formation.

The subset of 9 galaxies classified as LINERs frequently host both narrow, shocked
 components and broad, shocked components. In all but one of these objects, narrow,
shocked emission appears least 5 kpc from their associated nucleus. 
Among the 6 LINER spectra with broad, shocked components, the outflow
is spatially extended in 5 of them.
 None of the 9 galaxies have sufficient L$_{\rm [O III]}$ to suggest that the shocked
outflow or the narrow, shocked component is powered by an AGN NLR. These examples
emphasize that a integrated galaxy spectrum with a LINER classification need not
indicate the presence of an AGN.

Among the sample, 12 galaxies had emission line ratios on the borderline between 
classifications. 
 Six of these borderline galaxies have ratios between the Seyfert and LINER 
classifications.
 These six galaxies have broad outflows as well as extended shocks that are not part 
of the bulk outflow.  
 The two borderline galaxies with classifications between HII and LINER do not host 
shocked outflows, but do have extended shocks, which is responsible for making
the net classification slightly outside the range of the HII galaxies.
 the last of these twelve galaxies are a mix of classifications, without a clear trend 
in behavior.

The largest fraction of the sample (21 out of 48 galaxies) is classified as HII in the 
integrated spectra.
 These galaxies share the same HII classification in the diagnostic diagrams that 
compare [\ion{O}{1}]/Ha\ and [\ion{S}{2}]/\Ha, but the [\ion{N}{2}]/\Ha\ can be either
HII or ``composite'' as described in \cite{Kewley:2006p38}.
 The subset of galaxies with a ``composite'' classification are also host to either 
shocked outflows or shocks in the extended narrow emission line gas.
 Clearly, the presence of shocks in these objects modifies the [\ion{N}{2}]/\Ha\ 
classification, but observations with kpc scale spatial resolution showed that 
supernovae are a plausible source of the forbidden line enhancement and the outflow.

The spatially resolved analysis was also aided by the simultaneous 
multiple component fit to the emission line profiles, so we next  
investigate the integrated spectrum with the multiple component fit.  

\subsubsection{Identifying Shocks in the Integrated Spectral Profiles}
\label{sect:int_spect_prof}

\begin{figure*}[h]
\centering
\begin{tabular}{cc}
\epsfig{file=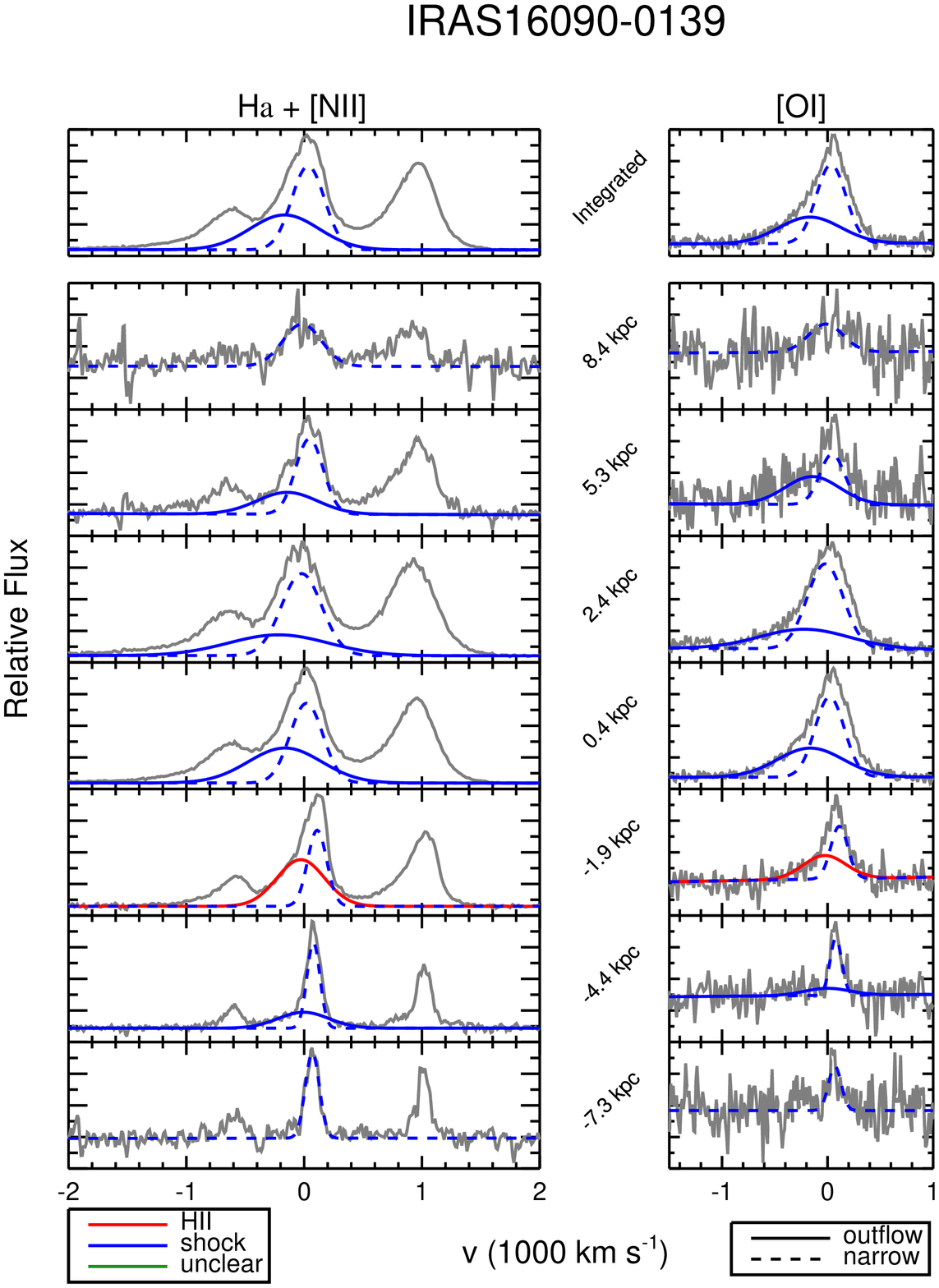, width=0.5\linewidth, keepaspectratio=true}
\epsfig{file=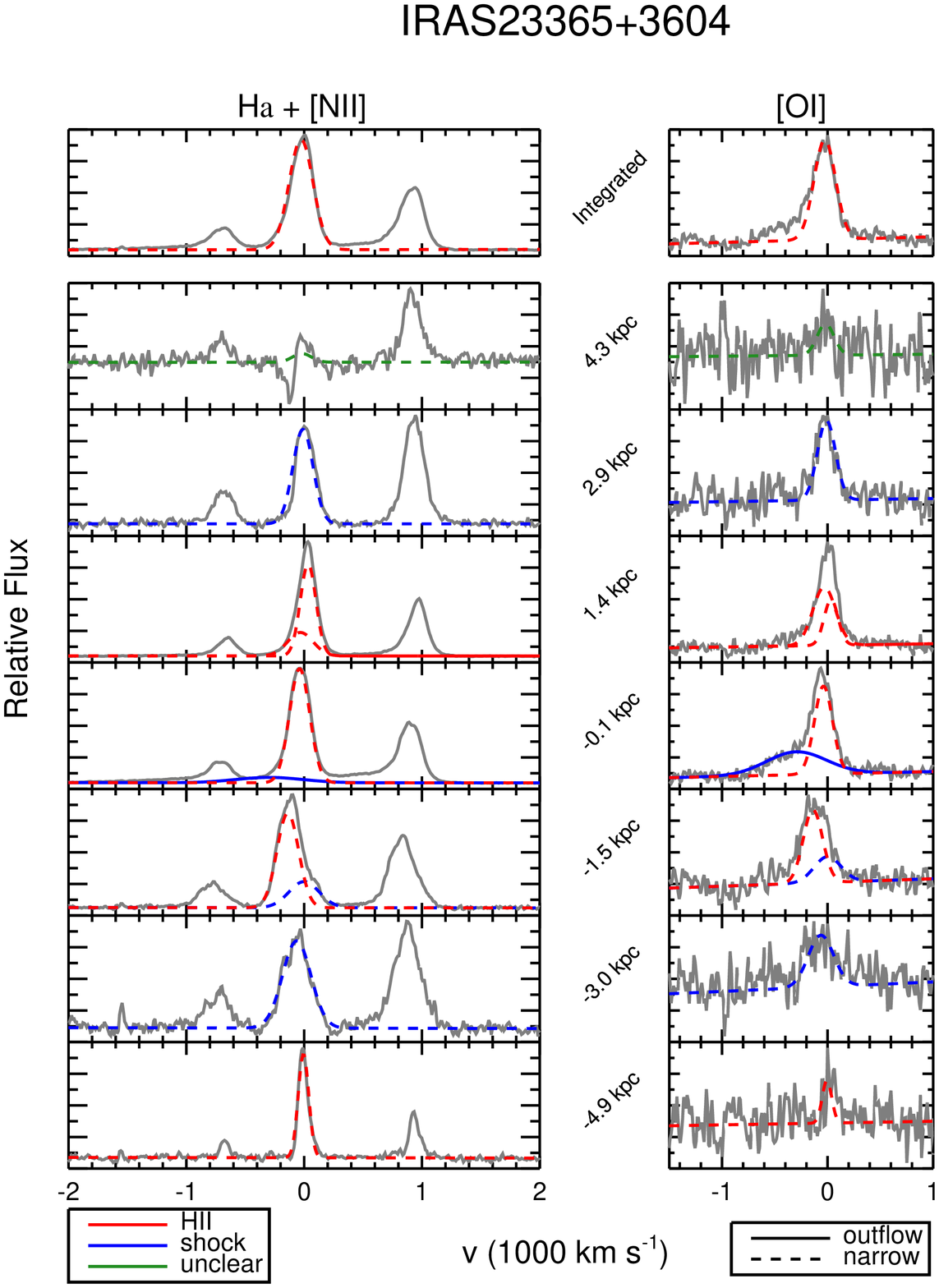, width=0.5\linewidth, keepaspectratio=true}
\end{tabular}
\caption{\label{fig:integ} Plot of the \Ha+[\ion{N}{2}] and [\ion{O}{1}] regions of the 
spectrum for both the integrated apertures and the spatially resolved apertures for 
IRAS16090-0139 and IRAS23365+3604. Line color represents the excitation classification of 
each component into ``HII'' (red), ``Shock-like'' (blue), ``unclear'' (green). The line style
identified the components that are part of an identified outflow (solid), compared to the rest 
of the emission components (dashed). {\emph Left:} IRAS16090-0139 presents outflow 
components in the resolved analysis that also appears in the integrated spectrum. The 
observed narrow shock also appears in the integrated profile fit. {\emph Right:} In the 
nuclear aperture of IRAS23365+3604 we identify outflow from the kinematics in the emission 
component. The apertures at 2.9 kpc and 3 kpc from the continuum source identify emission 
from shocked gas. All of these features are washed out in the integrated profile.}
\end{figure*}

Our analysis of spatially and spectrally resolved line emission provides new insight
on how to identify regions of shocked gas. Here, we examine whether the method
works on integrated spectra with moderate spectral resolution. 
This analysis differs from the previous section by comparing the line shapes
in the integrated profiles to those in the apertures.

This analysis allows us to define the required ranges and resolutions of 
instruments obtaining the integrated spectra.
Clearly, the integrated spectra must cover
[\ion{O}{1}]$\lambda6300$. The strong [\ion{O}{1}]$\lambda6300$ emission from shocked 
regions and the separation of this transition from other lines make it critical
for uniquely fitting multiple velocity components. Furthermore, we will assume
the integrated spectra have spectral resolution of at least $R \ge 5000$ 
and signal to noise $\sim 15$
if multi-component fitting will be attempted.

Of the 24 galaxies that exhibit broad blueshifted outflow in the spatially resolved 
analysis, 19 galaxies also show these features in the integrated spectrum.
In the remaining 5 galaxies, the strong narrow lines observed in the resolved 
analysis mask the signature of broad outflow in the integrated spectrum. 

Regions with narrow ($\sigma_v < 150$ \kms) emission lines that have shock-like line ratios 
appear frequently (34 of 48 galaxies) in the spatially resolved analysis.
Over a large range in position, these narrow features are identified at Doppler shifts 
ranging from -250\kms to 250\kms  in a few galaxies. Averaging over a large range in 
Doppler shift acts to broaden the integrated line profiles. In 5 of the 34 galaxies with 
narrow shock emission,  the summed components form a line of width $\sigma_v > 150$ \kms\ 
in the integrated spectrum.  These galaxies are mostly classified as HII 
with flux ratios in the ``composite'' region of the [\ion{N}{2}]/\Ha\ vs [\ion{O}{3}]/\Hb\ 
diagnostic diagram. Fig. \ref{fig:integ} shows the integrated aperture and spatially resolved 
apertures for IRAS16090-0139, which exhibits the outflow and narrow shocks from the resolved 
analysis in the integrated profile. We also compare this to the case of IRAS23365+3604, which 
shows both types of shock in the spatially resolved analysis, while the integrated case is 
only identified as HII.

In 15 out of the 34 galaxies, the Doppler shifts of the narrow, 
shocked components span a small enough range to be detected as 'narrow, shocked' components
in the integrated spectrum. A narrow span of Doppler shifts would also make 
any broad outflow more identifiable, which means that objects with low inclination gas disks
will show broad outflows more often.

We conclude that applying the multiple component fitting technique to 
integrated spectra would have some success in identifying broad, shocked 
 components from outflows. In our sample, the outflow is undetected in the integrated
spectrum only  20\% of the time.  In these 5 failures, a broad component is in 
fact detected in the integrated spectrum; but the spatially resolved analysis 
indicates the emission comes from a gas  rotating like a disk. Building a better
understanding of the physical origin of the narrow, shock-like emission
component is therefore critical for interpreting the presence of broad, shocked
emission in integrated spectra.

\subsection{Lessons From the Local Laboratories}
\label{sect:local_labs}

While there are important differences between ULIRGs and star-forming disks at $z>2$, ULIRGs 
present high specific star formation rates that are similar to the high end of 
starforming galaxies from 2.5~$<~z~<$~5 
\citep{Elbaz:2007p1052,DaCunha:2010p1071,Feulner:2005p1081}.
 The similar specific star formation rates suggest that local ULIRGs can provide 
insight into processes that occur below current resolution limits and that the 
broad emission components may share a common origin. 
 Using adaptive optics (AO) spectroscopy of $z \sim 2$ galaxies, star-forming clumps 
were recently shown to be the origin of the broad \Ha/[\ion{N}{2}] line wings
\citep{Genzel:2011p1007}. 
 The $\sim 4$ kpc offset of these clumps from the galactic center suggests a 
supernova-driven outflow, rather than an AGN, broadens the line emission. 
 The more central location of the broad emission found in the ULIRGs illustrates 
that the ULIRGs are not direct analogs of the $z \sim 2$ galaxies but the 
radius and luminosity of the broad emission region in most of these ULIRGs 
is uncomfortably large to be caused by the AGN.

In the echellete spectra, broad emission features are identified in the line profiles through the
simultaneous fitting of multiple transitions. 
 Our fitted linewidths are not as large as the broad \Ha/[\ion{N}{2}] wings reported
for $z \sim 2$ galaxies \citep{Shapiro:2009p867} because our spectra require 
broad components in the forbidden lines as well as the recombination lines.
 Additionally, since ULIRGs are known to be much more reddened than the $z \sim 2$ 
starbursts; broad emission-line profiles produced by similar physical 
processes in the less-dusty $z \sim 2$ galaxies will be more symmetric with
respect to the systemic velocity and therefore broader than those in ULIRGs. 
 While \cite{Shapiro:2009p867} had difficulty distinguishing a BLRs and superwinds
as a source of the broad emission features identified in stacks of $z \sim 2$ spectra,
the simultaneous fitting of more broad forbidden lines allows the identification of
other sources of the broad emission rather than just \Ha.

The estimates of outflow mass and outflow rate in the local ULIRGs provides context to the 
estimates of outflow rate and mass in the starforming clumps at $z \sim 2$ 
\citep{Genzel:2011p1007}. 
 The star formation rate in these local ULIRGs (24 - 180 \SFR) is 
much greater than the star formation rate in the $z \sim 2$ starforming clumps 
(3.3 to 40 \SFR), but our estimates of the mass outflow rate are much lower
(0.4 to 58 \SFR\ for ULIRGs, 6 to 200~\SFR\ for starforming clumps). 
These results are consistent with a higher $\eta$ for warm ionized gas 
at high redshift but do not demand it.  The gas density or the filling factor 
adopted for the high redshift galaxies could be tuned to give results 
for $\eta$ similar to what we infer for these ULIRGs.

Some properties of these outflows are easier to study in nearby galaxies, and the methods in this 
study illustrates how more information about the outflows at $z \sim 2$ can be obtained, even
in integrated spectra. 
 As mentioned already, for example, the outflows in $z \sim 2$ galaxies likely have
shock-like line ratios similar to those in our ULIRG sample.
 The properties of these shocks can be estimated once spectral coverage of additional
lines -- particularly [\ion{O}{1}], \Hb, and [\ion{O}{3}] -- 
are obtained at moderate spectral resolution, $R \approx 5000$, with similar signal to noise 
($\sim 15$) as in this study.

\section{Summary}
\label{sect:conclusions}

Paper~I mapped optical, emission-line ratios across 39 ultraluminous infrared galaxies
in 2 dimensions, i.e., line-of-sight velocity plus one spatial dimension. In this paper, we have
used those measurements to describe the kinematics of 
regions excited by different physical processes. Figure~2 shows the distribution of line-of-sight 
velocity dispersion for regions with shock-like and HII-like line ratios. The median linewidth
for the shock-like components, $\sigma_v \approx$ 144 \kms, 
is higher than that of the HII-like components, 61 \kms.
The difference is caused by the large number of shock-like components broader than 
$\sigma_v \approx 150$ \kms and the near absence of HII-like components with linewidths this large.

These broad emission components are relevant to our understanding of 
gas inflows and outflows because gas in virial equilibrium would produce lines
with smaller widths.
Our results provide insight into the interpretation of the emission-line spectra
of high redshift galaxies, particularly dusty galaxies with extremely high star formation 
rates.

We find that the broad shock-like components are typically spatially extended, reaching
radii up to 6~kpc in Figure~4. We show that in 9 out of 24 broad components, the
size of this region exceeds that expected for an AGN NLR under the 
assumption that [OIII] 5007 luminosity from components with log([\ion{O}{3}]/\Hb) $>$ 0.5 
is powered by the AGN. We estimate
the total power radiated by summing the optical luminosities of the broad line 
components and applying a geometrical correction for slit losses. Since the resulting
shock luminosities range from 5-20\% of the mechanical power from supernova explosions,
we conclude that supernovae are a viable power source for the gas flows that generate
the shocks.

In ultraluminous {\it infrared} galaxies, dust absorbs much of the starlight;
and the centers of these galaxies are not transparent at ultraviolet and optical
wavelengths. At least in the central few kpc, we can be reasonably certain that
the emergent emission-line profile is shaped by gas on the near side of the galaxy.
We therefore interpret the blueshifts of the broad components as direct evidence that 
the gas is outflowing. In less dusty galaxies, this outflow component would presumably 
also have a red wing from emission coming from the far side of the galaxy.

The broad components have surprisingly smooth line profiles. Across luminous
infrared galaxies with winds, for example, the \Ha\ + [NII] profiles are well described by a 
single, Gaussian component where emission from HII regions in the underlying disk
dominate the flux. When the outflows are observed against the sky, along the minor
axis, the emission lines are typically double peaked \citep{Lehnert:1995p1023}.
The relative intensities of the two components have been shown to reflect the
inclination and opening angle of a biconical outflow \citep{Heckman:1990p710}. That some
of the broad emission is detected beyond the continuum emission in these ULIRGs,
yet does not have a double-peaked profile may be related in part to the obscuration of the 
back side of the outflow. However, it is not obvious that the galaxies are opaque
at these large radii, and we tenatively conclude that the warm-ionized, outflowing
gas in ULIRGs does not share the biconical structure that describes outflows in LIRGs
very well.

The prominence of shock-excited emission in these starburst galaxies raises questions 
about how shocked emission skews the emission-line ratios measured from integrated 
spectra. Since the luminosity in the ultraviolet, ionizing continuum exceeds the 
mechanical power from supernova and stellar winds by about an order of magnitude 
\citep{Leitherer:1999p1117,Martin:2007p1080}, we would not expect shocks from galactic
winds to determine the line ratios measured in integrated spectra. What Figure~8
clearly demonstrates, however, is that when integrated spectra have moderately high
spectral resolution, $R \sim 5000$, shocked, outflowing gas can sometimes be recognized by comparing the
line profiles of forbidden lines, particularly [OI] 6300, to the blended \Ha\ + [NII]
profile. We conclude that multi-component line fitting of at least these 4 transitions
in the rest-frame optical spectrum can provide useful diagnostics of shock velocities
and radiative energy losses in galaxies over a very broad range of cosmic time.

\acknowledgments{The authors thank Kristian Finlator, Nicolas Bouch\'{e}, Alaina Henry, 
Vardha Bennert, Tommaso Treu, and Omer Blaes for illuminating discussions. 
This work was supported by the National Science Foundation under contracts AST-0909182 and 
AST-1109288 and the Department of Education through the Graduate Assistance in Areas of National Need 
program. A portion of this work was completed at the Aspen Center for Physics.
The authors wish to recognize and acknowledge 
the very significant cultural role and reverence that the summit of Mauna Kea has always had within the 
the indigenous Hawaiian community. We are most fortunate to have the opportunity to conduct 
observations from this mountain.}

{\it Facilities:}  \facility{Keck}

\bibliographystyle{apj}
\bibliography{ulirg2}

\end{document}